\documentclass{emulateapj}

\usepackage{epsfig}

\def\deg{\hbox{$^\circ$}}

\def\lesssim{\mathrel{\hbox{\rlap{\hbox{\lower4pt\hbox{$\sim$}}}\hbox{$<$}}}}
\def\gtrsim{\mathrel{\hbox{\rlap{\hbox{\lower4pt\hbox{$\sim$}}}\hbox{$>$}}}}

\def\beginrefer{\section*{References}%
\begin{quotation}\mbox{}\par}
\def\refer#1\par{{\setlength{\parindent}{-\leftmargin}\indent#1\par}}
\def\endrefer{\end{quotation}}
\shorttitle{Searching for B-fields in LSS}
\shortauthors{Xu et al.}
\begin{document}

\title{\bf A Faraday Rotation Search for Magnetic Fields in Large
Scale Structure}

\author{Yongzhong Xu\altaffilmark{1}, Philipp P. Kronberg$^2$, Salman
Habib$^1$, 
and Quentin W. Dufton$^2$}

\affil{$^1$The University of California, Los Alamos National
Laboratory, MS B285, Los Alamos NM 87545}

\affil{$^2$The University of California, Los Alamos National
Laboratory, MS C305, Los Alamos NM 87545}

\email{xuyz@lanl.gov, kronberg@lanl.gov, habib@lanl.gov, dufton@lanl.gov}

\begin{abstract}

Faraday rotation of radio source polarization provides a measure of
the integrated magnetic field along the observational lines of
sight. We compare a new, large sample of Faraday rotation measures
(RMs) of polarized extragalactic sources with galaxy counts in
Hercules and Perseus-Pisces, two nearby superclusters. We find that
the average of RMs in these two supercluster areas are larger than in
control areas in the same galactic latitude range. This is the first
RM detection of magnetic fields that pervade a supercluster volume, in
which case the fields are at least partially coherent over several
megaparsecs.  Even the most conservative interpretation of our
observations, according to which Milky Way RM variations mimic the
background supercluster galaxy overdensities, puts constraints on the
IGM magneto-ionic ``strength'' in these two superclusters. We obtain
an approximate typical upper limit on the field strength of about
$\sim 0.3~\mu$G ${l}/$(500~kpc), when we combine our RM data
with fiducial estimates of electron density from the environments of
giant radio galaxies, and of the warm-hot intergalactic medium (WHIM).
\end{abstract}


\section{Introduction}
\label{IntroSec}

The detection of a magnetized intergalactic medium (IGM) at levels of
a few~$\mu$G within clusters of galaxies (Kim et~al. 1990, Kim
et~al. 1991, Feretti et~al. 1995, Clarke et~al. 2001) raises the
question of what IGM magnetic fields might be detectable in the wider
intergalactic medium, well beyond galaxy clusters themselves, but
within galaxy overdense ``filaments'' of large scale structure
(LSS). The recent, much improved definition of filament-like galaxy
distributions in the Sloan Digital Sky Survey
(SDSS)\citep{aba03,aba04,aba05}, in addition to the more nearby Two
Micron All Sky Survey (2MASS)\citep{2MASS03} and the second Center for
Astrophysics galaxy redshift survey (CfA2)\citep{CfA98} at
$z{\lesssim}0.17$ makes it easier to identify galaxy overdense
filaments. In the local Universe, to which the two latter surveys
apply, the superclusters' larger angular extent encompasses an
optimally large number of radio sources with measured Faraday
rotations.

This paper describes an attempt, using the best RM data so far
available, to detect an RM signal due to an associated intergalactic
magneto-ionic medium in the Hercules, Virgo, and Perseus-Pisces
superclusters of galaxies. As discussed below, different theoretical
arguments and simulations cover a wide range of predictions, thus
strongly motivating the task of obtaining possible constraints from
direct observations.

\section{Arguments for and observations in support of a widespread
intergalactic magneto-plasma} 
\label{IGMSec}

\subsection{Proposed early magnetic field generation and amplification
in the course of LSS evolution}

Magnetic field seeding in the primeval intergalactic medium has been
proposed by Harrison (1970), in which chaotic motions of charged
particles can seed an initial magnetic field by a Biermann
Battery-like process \citep{Biermann50}. More recent calculations by
Gnedin, Ferrara \& Zweibel (2000) proposed to explain how the breakout
of ionization fronts from protogalaxies and their propagation through
hypothesized dense neutral filaments in the evolving web of large
scale structure can also generate, and amplify an intergalactic
magnetic field before $z\sim 5$. Essentially $\langle\partial
|B|/\partial t\rangle$ in the magnetic induction equation is positive,
so that weak pre-stellar intergalactic fields in effect convert
kinetic into magnetic energy. An IGM field during and after formation
of the first galaxies is a natural consequence of early galaxy
star-driven outflow, as modeled by Kronberg, Lesch, \& Hopp (1999),
especially earlier than $z\sim 7$. Analogous models by Furlanetto \&
Loeb (2001) for black hole/jet driven outflow have been shown to have
a similar IGM magnetic field seeding effect at earlier epochs.

Space-filling fields at early epochs can be considered as seed fields
to incorporate into computational models of large scale structure
(LSS) evolution.  The intergalactic fields could be amplified in the
course of gravitational collapse of baryonic and dark matter
into the ``filaments'' of large scale structure, essentially
converting gravitational infall energy into magnetic energy. The
combination of shearing and shocks have been calculated by some
authors (e.g., Ryu et al. 1998) to generate magnetic fields up to
$\sim$ 0.1~$\mu $G in these intergalactic filaments. Up to now, there
has been no observational verification of the actual field strengths
in large scales in filaments. Indeed, independent LSS evolution
simulations by other groups (e.g., Dolag et al. 2005) arrive at very
low LSS IGM magnetic field strengths, too weak to be detectable in our
present Faraday rotation data. Remaining uncertainties in the detailed
physics of these difficult simulations underline the importance of
direct observational tests, one of which we describe in this paper.

\subsection{Black hole seeded intergalactic magnetic fields and cosmic
rays}

A very different source of IGM magnetization originates in the
reservoir of gravitational infall energy in galactic central black
holes (e.g., Kronberg et al. 2001). The average comoving density of
galactic black holes (BH) in the mature universe is estimated at
$\rho_{BH}\sim 2\times 10^{5}$~M$_\odot$/Mpc$^{3}$ and they are
widespread in the population of galaxies (e.g., Gebhardt et
al. 2000). When we combine this fact with the radio source evidence
that central BH - driven jets and radio lobes convert a substantial
fraction of the BH formation energy into magnetic fields that are
transferred to the IGM via radio lobes (e.g., Kronberg et al. 2001),
the BH-generated magnetic field strengths in galaxy walls and
filaments is estimated to be of order 0.1~$\mu $G or more (Colgate and
Li 2004).

Independent calculations of the BH photon energy release during the
quasar epoch near $z=2$ have been made by Soltan (1982), Small \&
Blandford (1992), and Chokhsi \& Turner (1992). The efficiency of
conversion of the presumed central BH infall energy to photon energy
released is remarkably high, a few tens of percent.  Analyses of giant
radio galaxies (GRGs) show that the conversion efficiency of this BH
infall energy into magnetic energy is comparably high (Kronberg et
al. 2001, 2004), i.e. the magnetic energy, cosmic ray energy and
photon energy released into the IGM are comparable.

\subsection{Estimates of a detectable LSS RM signal based on analyses
of Mpc scale BH-powered radio sources} 
\label{PossRMSec}


The Faraday rotation through an intergalactic region permeated by a
magnetic field and thermal (non-relativistic) gas is
\begin{equation}
\label{RMEq}
\hbox{RM} = 0.81 \cdot
\frac{n_{th}^{IG}}{\hbox{cm}^{-3}}\cdot \frac{B_{{\|}}}{\mu\hbox{G}}
\cdot {\frac{L}{\hbox{pc}}}~\hbox{rad m$^{-2}$}.
\end{equation}
Over a dimension $L=1$~Mpc, containing a 1~$\mu$G field and having
$n_{th}^{IG}=10^{-6}$~cm$^{-3}$, the estimated RM $\sim
1$~rad~m$^{-2}$. The RM would be reduced by ${\cal{O}}(N^{1/2})$ where
$N$ is the number of field reversals over a line of sight through the
supercluster. The outer lobes of GRGs ($\gtrsim$1~Mpc) contain at least
partially aligned fields over at least a fraction of a megaparsec
(e.g., Kronberg, Wielebinski \& Graham 1986, Mack et al.
1997).  Similar evidence is provided by the peripheral regions of the
synchrotron-emitting cosmic ray halos of some galaxy clusters (Clarke
\& En{\ss}lin 2005). These facts suggest that some early component of large
scale ordering existed within typical supercluster volumes.

Information on the plasma densities from direct observation can be
gleaned from detailed, multifrequency observations of a few GRGs
having dimensions up to $\sim 4$~Mpc. Several such systems have
established RM variations that probe the ``ambient'' intergalactic
medium within $\sim 3$~Mpc of the radio source and give estimates of
$n_{th}^{IG} |B^{IG}|$. Typically, $n_{th}^{IG}\simeq
10^{-6}$-$10^{-5}$~cm$^{-3}$ for the ambient IGM on $\sim$1-3~Mpc
scales around the GRGs, and magnetic field strengths, based on the
assumption of particle/field energy equipartition are close to
1~$\mu$G (e.g. Willis \& Strom 1978, Strom \& Willis 1980, Kronberg,
Wielebinski and Graham 1986). Magnetic field strengths at these levels
are independently consistent with values of $\lesssim 1~\mu$G
estimated by Kim et al. (1989) from very low levels of intergalactic
synchrotron emission at 0.3~GHz that they detected in the IGM out to
$\sim 2$~Mpc from the Coma cluster of galaxies.

The above measurements lead us to estimate possible RM$_{igm}$ values
from a supercluster environment that could be as high as
$\sim$20~rad~m$^{-2}$ in an optimistic scenario, for a typical
supercluster pathlength scale of $\sim5$~Mpc, $n_{th}^{IG}\sim 5\times
10^{-6}$~cm$^{-3}$, and $\langle |B_{IG}| \rangle \sim 1$~$\mu$G. It
is more likely that RM contributions from LSS are typically less than
$\sim $ 20~rad~m$^{-2}$, but some fraction of this RM
could be detectable, depending on our ability to remove Galactic
foreground RM, and this is what motivates us to attempt it.

\section{Data and analysis tools}
\label{DataSec}

\subsection{Faraday Rotation Measure Data}
\label{RMDataSub}

We have used an augmented version of the extragalactic source Faraday
rotation measures published by Simard-Normandin, Kronberg and Button
(1981). The number of RMs that are suitable for our LSS filament
probes is roughly double that in the Simard-Normandin, Kronberg \&
Button catalog of 1981. The accuracy of some of the RMs has also been
improved in the present analysis by the addition of more radio
polarization measurements. We omitted the very few RM sources that are
in front of the local superclusters so as to have only background RM
``illuminators'' of the supercluster medium.

We define the supercluster boundaries using optical galaxy counts as
explained further below. Because Faraday rotations of radio galaxies
and quasars at $z\gtrsim 0.5$ are commonly affected by optically
unseen intervening systems and source-intrinsic evolutionary effects
(Kronberg \& Perry 1982, Welter, Perry \& Kronberg 1984, Athreya et
al. 1998) we have omitted the RMs of most sources that are at large
redshift since they, as expected, show RM anomalies that are unrelated
to the RM-{\em weak}, local universe intervenors that that we wish to
probe. We also removed RMs of all sources lying within 7$^{\circ}$ of
the Galactic plane because of the strong RM perturbations due to the
Milky Way. Finally, since galaxy clusters are known to perturb the
Faraday rotations of radio sources within and behind them (Kim et al.
1990, 1991, Clarke et al. 2001), we have removed RM sources within
0.5$^{\circ}$ of known galaxy clusters.

The increased numbers of Faraday RMs now available to us are
sufficiently large that we have for the first time, relatively good
sky coverage behind and beside the local superclusters. We used two RM
subsets, (i) sources with known $z \leq 0.5$, and (ii) a comparable
number of sources with unknown $z$, but from which we have rejected a
small number of anomalous values which could be due to high $z$
effects or intermediate redshift clusters. The second sample is
comparable in size to the first, and gives us substantially more
sources at the expense of some possible high-$z$ effect
``pollution". Consequently we combined the two subsets in all
subsequent analysis, and have ``cleaned" out statistically deviant
high RMs.

We distinguish between two kinds of RMs: the individual measured source RMs and
the smoothed RM (SRM). The latter are a combination of galactic features and,
possibly, given the ``expectation'' arguments outlined above, degree-scale
contributions from local-universe large scale structure. SRMs are obtained by
averaging individual RMs within a circle of a radius of $15\deg$, centered on
the sources. The SRM smooths out small-scale RM fluctuations due to the radio
sources, and should mainly contain information on magnetic fields from large
angular scale structures. In order to display the SRM variations more clearly
and to compare with the superclusters visually, we Gaussian-smooth the SRMs to
a 7-degree scale, to match the smoothing scale for the 2MASS data; A 7-degree
smoothing is sufficient to clearly display the supercluster profiles. The
upper panel of Fig.~1 shows a plot of the Gaussian-smoothed distribution of the
radio source SRMs for the galactic sky that are relevant to the present
analysis. The black points show the positions of the individual sources used in
the SRM calculation. The colorbar gives the scale for our best estimate of the
SRM over the sky.  In order to more clearly illustrate the contrast of the
lower SRMs, we saturated the SRM color scale at $-65$~rad~m$^{-2}$ and
$+65$~rad~m$^{-2}$ in blue or red respectively. The larger SRMs extended to
$\sim$ $|100|$ rad~m$^{-2}$. The southern sky below $\delta = $-25$^{\circ}$ is
excluded because of the lower angular density of RMs, hence less
well-determined SRMs. We also exclude the area near the galactic equator, i.
e., $|b|\leq 7^{\circ}$, due to the strong galactic contribution to SRM
\citep{SK80}.

\begin{figure*}
\label{SRMFig}
\plotone{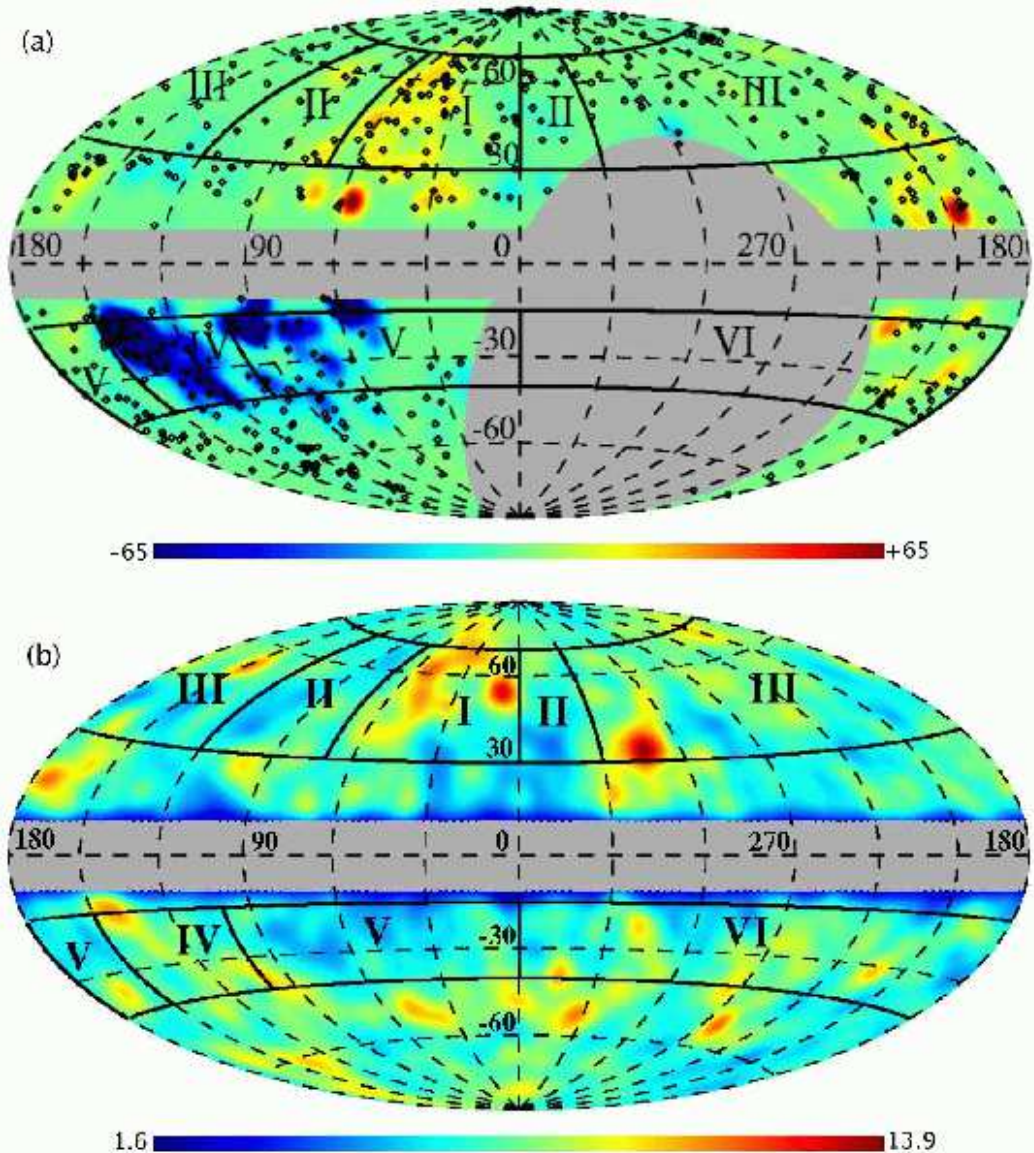}

\caption{(a) All-sky plot of the extragalactic source SRM
distribution. The square black points show the positions of individual
SRM sources. The color map shows SRM variations Gaussian-smoothed to 7
degrees in an Aitoff projection of galactic coordinates. The color
range from blue to red signifies a range of SRM values from
$-65$~rad~m$^{-2}$ to $65$~rad~m$^{-2}$. The excluded areas (see text)
are for $|b|$ $\leq $7$^{\circ}$ and $\delta \leq $-25$^{\circ}$ (b)
The 2MASS galaxy column density Gaussian-smoothed to 7 degrees in the
same coordinates as in (a). The Hercules and Perseus-Pisces
superclusters lie in the high density parts labeled I and IV,
respectively.  Labels II, III and V indicate the RM comparison zones
(see subsection~\ref{RMDataSub}, and table~\ref{CompaTab}).}
\vspace{.5cm}
\end{figure*}

\subsection{The CfA2 and 2MASS Galaxy Surveys}

We have utilized two sets of galaxy survey data for our analysis, the
second Center for Astrophysics survey and the Two Micron All Sky
Survey.

The second Center for Astrophysics (CfA2) Redshift Survey covers the
whole northern celestial hemisphere, and has measured redshifts of
about 18,000 bright galaxies with apparent blue magnitude $m_b$ less
than 15.5. This survey captures the three-dimensional properties of
the large-scale structures in the nearby universe ($z<0.05$), which is
very useful for RM-LSS cross-correlation analysis~\citep{CfA98}.

The Two Micron All Sky Survey (2MASS) has covered 99.998\% of the sky
observed from both northern and southern 2MASS observation sites
\citep{2MASS03}. The 2MASS has three infrared cameras in the J
($1.25$~$\mu$m), H ($1.65$~$\mu$m), and Ks ($2.17$~$\mu$m)
bands. Near-infrared observations are useful for lines of sight near
the Galactic Plane since the extinction due to the interstellar medium
at near-infrared wavelengths is only about one-tenth of that at
optical wavelengths. The all-sky data release from 2MASS includes
470,992,970 point sources and 1,647,599 extended sources or galaxies
in the nearby universe. In order to keep the data uniform, we only use
those extended sources with apparent Ks-band magnitude brighter than
14.5 (i.e., $m_{Ks} < 14.5$).  This still leaves 1,312,603 extended
sources for our analysis.

The 2MASS contains only two-dimensional information of LSS, while the
CfA2 contains three-dimensional information. Therefore, as described
in Subsections~\ref{2MassSub} and \ref{CfA2Sub} in detail, we use
different tools to analyze the data from the two surveys. For 2MASS,
we compare the column density with RMs, calculated using the HEALPix
algorithm (section 4.1), whereas for CfA2, we use the Voronoi
tessellation method, described next.

\subsection{3D Voronoi Tessellation}

\label{VTSub}

Voronoi Tessellation (VT) is a robust tool for calculating number
densities in two-, three- or higher-dimensional spaces. Here, we use a
very efficient VT algorithm, QHULL (Barber, Dobkin \& Huhdanpaa 1996),
to calculate the galaxy number density in the CfA2 catalog. The
procedure is summarized below.

First, using QHULL, we construct the entire volume of the CfA2 catalog
with a number of Voronoi Diagrams (VDs). Each VD corresponds to a
galaxy in this catalog. The VD of a given galaxy is a convex hull
bounded by a series of planes that bisect the segment between this
galaxy and its neighbors. In consequence, any query point inside this
VD is closer to this galaxy than any other galaxy. The ``proprietary
volume'' of a galaxy is that of its VD \citep{Icke87}. The local
galaxy number density is the inverse of its volume, i.e., $\rho_i =
1/\hbox{Vol}_i$, where $\rho_i$ and $\hbox{Vol}_i$ are the number
density and volume for the $i^{th}$-galaxy. The overdensity of a
galaxy is defined as
\begin{equation}
\label{OverDenEq}
 \delta_i = (\rho_i -\bar{\rho_i})/\bar{\rho_i},
\end{equation}
where $\bar{\rho_i}$ is splined from $\bar{\rho}(z)$, which is the
mean number density at a given redshift bin at $z$.  The rotation
measure should relate to both the galaxy number density and path
length of a radio ray passing through a portion of large-scale
structure volume.  Due to the survey selection function, the mean
number density $\bar{\rho}$ depends strongly on galaxy redshift. In
order to eliminate the effect of this selection function, we employ an
estimator of the galaxy {\it overdensity} by effectively normalizing
out the survey selection function. The overdensity describes the
relative fluctuations of galaxy density, therefore in principle it
should not depend on redshift. At this point we are in a position to
use the overdensity of galaxies for our correlation analysis.

\subsection{Pathlength and weighted pathlength}
\label{PathSub}

Aside from the overdensity, another important parameter is the
pathlength of radio sources through the superclusters. First, we need
to ascertain whether the line of sight to a radio source passes
through superclusters. Second, we need to calculate the effective
pathlength. Since we have tessellated the superclusters into a large
number of VDs, the pathlength to a radio source is the sum of lengths
inside the VDs that radiation from the source passes through, i.e.,
\begin{equation}
\label{PathEq}
 pl_j = {\displaystyle\sum_i l_{ij}},
\end{equation}
where $pl_j$ is the pathlength, and $l_{ij}$ is the length inside the
$i^{th}$ VD which radiation from radio source $j$ passes through. We
also introduce a related parameter, the weighted pathlength, $wpl_j$,
which is the sum of the product of overdensity and length inside VDs:
\begin{equation}
\label{WPathEq}
 wpl_j = {\displaystyle\sum_i}l_{ij}\delta_{ij},
\end{equation}
where $\delta_{ij}$ is the overdensity of the $i^{th}$ VD which the
radio source $j$ passes through. As defined in Eqn.~\ref{OverDenEq},
overdensity is a linear function of the real galaxy number density,
Hence, the product of overdensity and pathlength is a linear
function of galaxy column density at the specific VD. 

\section{Results and Discussion}


In order to correlate the RMs of radio sources with galaxy overdensity
in nearby superclusters, it is crucial to find a way to identify those
superclusters or local filaments of large-scale structure. We have
developed a fast algorithm to identify superclusters after
tessellating all galaxies in the CfA2 catalog. The algorithm will be
described elsewhere~\citep{xu2005}. The basic idea is to find
structures (or continuous Voronoi Diagrams) with overdensities in a
certain range. For CfA2, we found that an overdensity larger than 2
sets a very reasonable threshold to identify superclusters. Using this
algorithm, we successfully identified and defined 9 nearby
supercluster candidates.

Out of 9 supercluster candidates that we initially identified in the
CfA2 catalog, only the Perseus-Pisces, Hercules, and Virgo
superclusters give a sufficiently large angular coverage in a
favorable part of the galactic sky, with a sufficiently good match of
galaxy column density and angular RM coverage on the sky. This is
needed to ensure that we have a sufficient number of radio sources in
those sky areas to statistically analyze the data. The Virgo
supercluster is the least favorable of these three, being
inconveniently close to us, hence distributed over a very large area
of sky, so that galactic foreground RM effects are more difficult to
remove. Further, the low angular number density of Virgo supercluster
galaxies is relatively more contaminated by background
galaxies. Therefore, we compare the RMs with 2MASS data only for the
Perseus-Pisces and Hercules superclusters, although we did attempt the
VD method for Virgo (Fig.~2).

In this analysis it is important to recall that ``foreground'' SRM
features due to the Milky Way probably exist at some level at the higher
galactic latitudes where our superclusters lie.
 Our search for an RM signal from local supercluster
fields can therefore only rely on detailed correlations between zones
of galaxy overdensity and of the SRM. The (reasonable) premise here is
that Milky Way-induced SRMs are not correlated with extragalactic
magnetized regions.  Even if galactic $+$ intergalactic superpositions
exist at weak RM levels we can, at minimum, set the first useful upper
limits to SRM contributions from supercluster zones. Such upper limits
are interestingly close to the predictions of theoretical
calculations, as discussed in Section~3.

There is an intriguing juxtaposition of a Milky Way and extragalactic
RM contribution in Region A, a well-known anomaly in the RM sky
(Simard-Normandin \& Kronberg 1980). Region A, the huge negative RM
zone located in $50\deg\leqslant l\leqslant 150\deg$ and $b\geqslant
-40\deg$ [see the large deep blue area in Fig.~1(a)], overlaps in part
with the Perseus-Pisces supercluster, but has a greater overall
angular extent. As we discuss below, there is a weak, but positive
correlation (Table~1) between the SRM and galaxy overdensity in this
zone. If this correlation is not being (perversely) masked by galactic
effects, it would indicate that ordered magnetic fields in the
Perseus-Pisces supercluster maintain the same prevailing sense on
megaparsec scales. Unfortunately firm conclusions on this question
cannot yet be drawn because of possible ``interference'' by the Milky
Way foreground RM. Alternatively stated, the P-P supercluster filament
``rides'' on the large negative RM zone of region A. One possible
interpretation is that the P-P supercluster contribution to the SRM
(Fig.~1) contains a positive RM component that is not strong enough to
overwhelm the galactic RM contribution in this same area.

The RM region A does not appear to correlate with other known large
scale features of the galactic sky, such as the HI column density,
diffuse galactic synchrotron emission, gamma rays, etc. The
galactic feature that {\it does} generally coincide
with region A is the diffuse 45$\mu$m radiation, due presumably to
galactic dust~\citep{SFD98}. It is further salutary that the RMs in
Region A are predominantly negative, which conforms to the prevailing
direction of the interstellar field in the Sun's vicinity. However
this fact by itself need not necessarily refute the existence of a
mixed galactic/extragalactic origin for the anomalous Region A.

\subsection{Correlation between RMs and 2MASS Data} 
\label{2MassSub}

2MASS is an all-sky photometric survey with three infrared cameras. An
easy way to calculate the 2MASS galaxy number density is to pixelize
the all-sky data appropriately. Pixels corresponding to the location
of large-scale structures should have a higher number density of
galaxies. The public software package Hierarchical Equal Area
isoLatitude Pixelization (HEALPix) of the sphere~\citep{GHW} is
well-suited to this purpose. We pixelize the 2MASS data into
$12\times128^2$ cells. Since each pixel has the same area, we need
concern ourselves only with the number of galaxies in each pixel. In
order to obtain a continuous density field, we smooth the galaxy
column density ($N_g$) onto a $1\deg$ scale to eliminate
discontinuities on very small scales. After smoothing, the mean galaxy
number in each pixel is around 7. Therefore, we classify the number
density into three ranges: an underdense range ($N_g < 7$), one sigma
higher than the mean density range ($7 \leqslant N_g \leqslant 10$),
and an overdensity range ($N_g > 10$).

\begin{table*}[th]
\begin{centering}
\caption{Comparison of the mean values of absolute SRM
in areas I, II, and I+II+III, and in areas IV and V.}
\label{CompaTab}
\begin{tabular}{c|cc}
\hline\hline

Sky Area & Galaxy Density Range & SRM mean \\
 & & [rad m$^{-2}$] \\

\hline\hline
& all & {$\bf 14.3\pm 2.0$}  \\
Hercules Area I, 50 sources  & $N_g < 7.$ & 14.8 \\ 
($ 0\deg \leqslant l \leqslant 70\deg $, $30\deg \leqslant b \leqslant
70\deg $) & $ 7. \leqslant N_g \leqslant 10.$ & 15.9\\  
& $N_g > 10.$ & 12.9  \\
\hline
 & all & {$\bf 10.0\pm 1.8$}  \\

Hercules Comparison Area II & $N_g < 7.$ & 8.7  \\ 
($70\deg \leqslant l \leqslant 120\deg $ or $l \geqslant 330\deg$,
$30\deg \leqslant b \leqslant 70\deg$), 30 sources  & $N_g \geqslant
7. $ & 10.9  \\  
\hline
 & all & {$\bf 10.2\pm 0.8$}  \\
Hercules Comparison Zone, Area I+II+III, 154 sources & $N_g < 7.$ &
11.4  \\ 
 ($0\deg \leqslant l \leqslant 360\deg $, $30\deg \leqslant b
\leqslant 70\deg$) & $ 7. \leqslant N_g \leqslant 10.$ & 8.7  
\\ 
& $N_g > 10.$ & 10.5  \\
\hline\hline
	& all & {$\bf 56.8\pm 8.6$} \\
Perseus-Pisces Area IV, 45 sources & $N_g < 7.$ & 65.7  \\
($100\deg\leqslant l \leqslant 150\deg$, $-40\deg\leqslant b \leqslant
-15\deg$) & 
$ 7. \leqslant N_g \leqslant 10.$ & 53.5  \\
	& $N_g > 10.$ & 55.5  \\
\hline
 & all & {$\bf 41.0\pm 6.3$}  \\
Perseus-Pisces Comparison Area V, 42 sources & $N_g < 7.$ & 49.4 \\
($150\deg\leqslant l \leqslant 180\deg$ or $ l\leqslant 100\deg$,
$-40\deg\leqslant b \leqslant -15\deg$)  
	& $ N_g \geqslant 7. $& 32.6 \\
\hline\hline
\end{tabular}

\tablecomments{As illustrated in Fig.~1, areas I and IV are high
galaxy density areas where the Hercules and Perseus-Pisces
superclusters are located respectively, and areas II and V are low
galaxy density areas adjacent to these two superclusters.}
\end{centering}
\vspace{.5cm}
\end{table*}

To illustrate the nearby superclusters more clearly, we
Gaussian-smooth $N_g$ on a scale of 7 degrees as shown in the lower
panel of Fig.~1.  The Hercules and Perseus-Pisces superclusters constitute 
the high density parts labeled I and IV, respectively. Labels II and
III indicate the (low galaxy density) RM comparison zones for region I
in the same Galactic latitude ranges ($30\deg \leqslant b \leqslant
70\deg $); Label V indicates the comparison zone of region IV
($-40\deg\leqslant b \leqslant -15\deg$). We do not use region VI
since the RM data there are less complete.

As estimated in Section~\ref{PossRMSec}, RM values from a supercluster
environment might be as high as $\sim$20~rad~m$^{-2}$. If we assume
that $N_g$ is proportional to the pathlength of radio sources through
large-scale structures, and that $n_{th}^{IG}$ is constant, we would
expect positive correlations between RMs and $N_g$. Due to the unknown
complexity of alignment and reversal scale of the IGM magnetic field,
it is hard to find a direct correlation between the absolute RMs and
galaxy counts in the corresponding pixels. However, we may expect that
the mean values of absolute RMs in a supercluster area should be
larger than in other areas without superclusters. Due to the magnetic
field of the Galaxy, however, observed SRMs strongly depend on the
galactic latitude \citep{SK80}. We therefore analyze the RM data in
two corresponding spherical rings (see Table~\ref{CompaTab} and
Fig.~1). Table~\ref{CompaTab} lists mean values of absolute SRM of the
various areas classified above. We have filtered out zones of
anomalously high local galaxy density, so as to avoid contamination by
e.g. galaxy clusters.

For the Hercules supercluster zone we find that over its {\it entire}
area the mean SRM is consistently larger than in the two comparison
zones. It is close to, but does not coincide with the North Galactic Spur (Haslam et al. 
1982). This systematic difference also holds for each subrange of
galaxy density (Table~1).  However, whereas there is a global
correlation of enhanced SRM with the overall boundary of the Hercules
supercluster, the detailed locations at 7${\deg}$ resolution of maxima
in smoothed galaxy column density with those of the SRM [comparing
Figs.~1(a) and (b)] do not correspond exactly. This is also reflected
in the plot of weighted pathlengths (Section~\ref{PathSub}) against RM in
Fig.~2, in which we used the VD method and the CfA2 data (Section~\ref{CfA2Sub}).
We can interpret the combination of these two results to
indicate that an RM component is generated in the Hercules supercluster, but if so,  
most ordered magnetic field regions avoid the zones of highest galaxy density. This
conclusion must remain tentative until we better understand the
galactic foreground RM (which may also play some role, as suggested by the galactic 
pulsar RMs). A future higher density of pulsar RMs is most crucially needed to clarify 
the decomposition of galactic and supercluster RM contributions in this high latitude zone. 

In the case of the Perseus-Pisces supercluster, the SRMs within the
overall supercluster boundary are similarly larger, again for all
galaxy density regimes (Table~1). In contrast to the Hercules
supercluster, there is also a suggestion of a positive correlation on
smaller scales between SRM and galaxy density. This is seen both in
Fig.~1, and in the plots (Fig.~2) of $wpl$ vs. SRM.

For the Virgo supercluster, we find no evidence for a correspondence
on any smoothed scale (Figs.~1 and 2).
       
\subsection{Correlation between RMs and CfA2 data}
\label{CfA2Sub}

\begin{figure*}[t]
\label{CorrFig}
\plotone{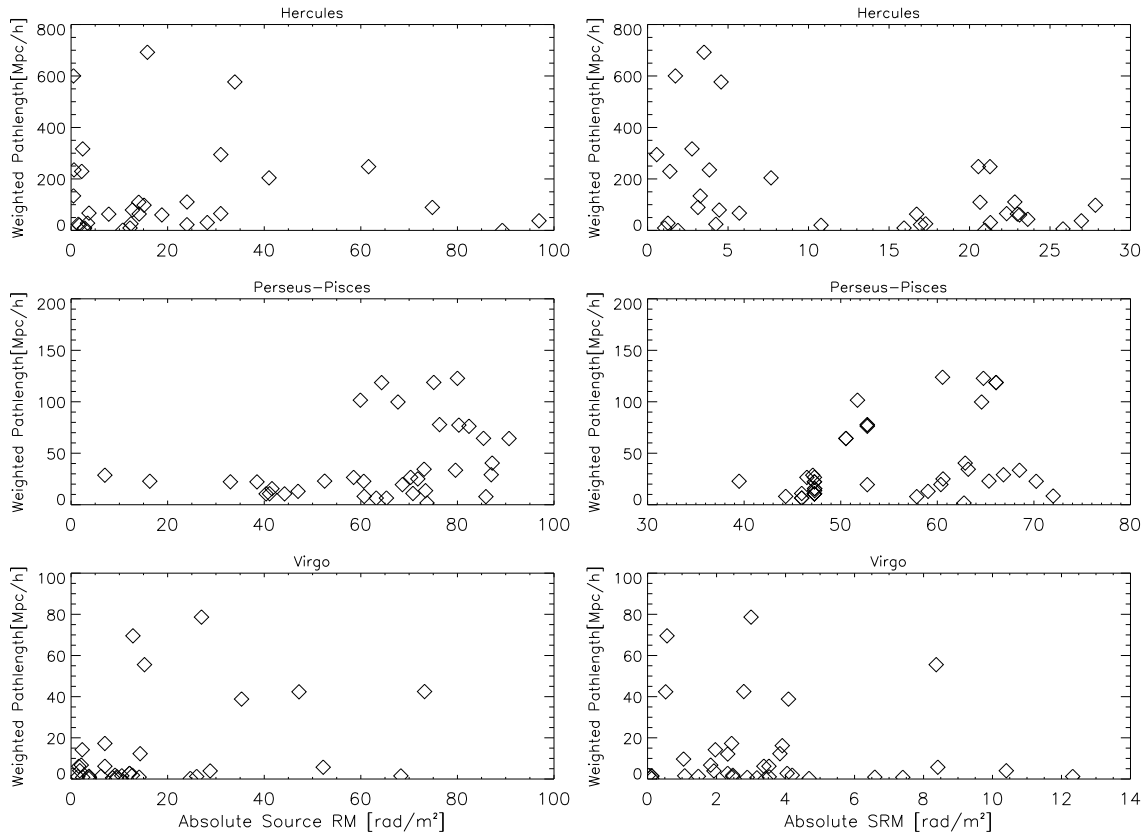}
\caption{Plots of weighted pathlength $wpl$ vs. source RM and SRM 
respectively, from our VD-based analysis of the CfA2 catalog. The left three panels, from top to bottom, show the
relation of $wpl$ and source RM in Hercules, Perseus-Pisces, and Virgo
superclusters respectively, and the right three panels show the
relation of $wpl$ and SRM in those three superclusters.}
\vspace{.5cm}
\end{figure*}

The CfA2 galaxy redshift survey covers the northern celestial
sky. Using the Voronoi Tessellation technique mentioned in
Section~\ref{VTSub} and in \citet{xu2005}, we identify galaxy
superclusters in the CfA2 catalog. Then, from
Eqns.~\ref{PathEq}-\ref{WPathEq}, we calculate the real pathlength
$pl_i$ and weighted pathlength $wpl_i$ of the $i$-th radio source
passing through the supercluster. Although $wpl_i$ has the dimension
of length, it is a linear function of galaxy number column density,
which we assume is proportional to the column density of thermal
electrons in the IGM. The RM is proportional to the electron column
density and the line of sight component of the magnetic field
(Cf. Eqn.~\ref{RMEq}). Therefore, we expect a positive correlation
between RMs and $wpl_i$. Fig.~2 shows plots of weighted pathlength
$wpl$ vs. source RM and SRM respectively. The left three panels, from
top to bottom, show the relation of $wpl$ and source RM in Hercules,
Perseus-Pisces, and Virgo superclusters respectively, and the right
three panels show the relation of $wpl$ and SRM in those three
superclusters. We find that the $wpl$ is positively correlated to the
fluctuations of absolute SRM in the Perseus-Pisces supercluster, but
that there are no such correlations in the other two superclusters
(see above).

\subsection{RM data vs. RM expectations in the Hercules supercluster} 
\label{RMExSub}

In Section~\ref{PossRMSec}, it was roughly estimated that the RM in a
supercluster may reach $\sim20$~rad~m$^{-2}$, scaled to $pl \sim 5$~Mpc,
$n_{th}^{IG}\sim 5\times 10^{-6}$~cm$^{-3}$, and $\langle|B_{IG}|\rangle\sim
1$~$\mu$G, by using the simple model described in Eqn.~\ref{RMEq} and the
related text in Section~\ref{PossRMSec}. RMs estimated from this
model are designated as MRMs. Comparing these MRMs with the measured SRMs, we
put constraints on the IGM magnetic fields and electron density. The detailed
procedure is described as follows. First, we use the simple model to estimate
MRMs of all radio sources in the Hercules supercluster for different  magnetic
field reversal scales by varying $n_{th}^{IG}$ and $|B_{IG}|$. Second, we calculate the
$\chi^2$ of each given pair of $n_{th}^{IG}$ and $|B_{IG}|$, for a given
$B$-field reversal scale. Specifically,
\begin{equation}
\chi^2 = \displaystyle\sum_{i=1}^{N_s}(MRM_i -
SRM_i)^2/\sigma_{SRM_i}^2,  
\end{equation}
where $N_s$ is the total number of radio sources passing through the Hercules
supercluster. Third, as usual, the corresponding normalized likelihood is
defined as $\mathcal{L}=\exp{(-(\chi^2-\chi_{min}^2)/2)}$, where the
$\chi_{min}^2$ is the minimum $\chi^2$ of a certain pair of $n_{th}^{IG}$ and
$|B_{IG}|$. Finally, we may derive the $1\sigma$, $2\sigma$, and $3\sigma$
contours from likelihood values equal to 0.607, 0.135, and 0.011 respectively
as shown in Fig.~3. From the calculated likelihood, we obtain constraints on
the magnetic field and the thermal electron density. The four panels in Fig.~3
show the constraints on $|B_{IG}|$ and $n_{th}^{IG}$ for magnetic field
reversal scales of 200, 400, 600, and 800~kpc respectively. The constraints have
the expected dependence on the reversal scales, the larger reversal scales
yielding a lower-magnetic-field constraint with a given electron density.  This
follows straightforwardly from Eqn.~\ref{RMEq}. Fig.~3 illustrates how the
$B$-field, reversal scale, and $n_{th}^{IG}$ are strongly degenerate. 
Independent $n_{th}^{IG}$ measurements are needed to help break this
degeneracy. For instance, the most recent warm-hot IGM (WHIM) observations
imply a hydrogen density of $\sim 5\times 10^{-6}$~cm$^{-3}$
\citep{Nic2005a,Nic2005b}. If we combine this global estimate with our results
here, we obtain $B$-field constraints of $0.4\pm0.2$~$\mu$G ($2\sigma$) and
$0.3\pm0.1$~$\mu$G ($2\sigma$) with reversal scales of 400~kpc and 800~kpc
respectively.  Here, we use the SRMs as distinct from the RMs generated by the
medium in the filaments. Since the SRMs may contain a Milky Way component,  the
absolute values of SRMs are statistically larger than the RMs generated in the
filaments.  In this sense, the constraints shown in Fig.~3  should be
considered as upper limits for the magnetic fields in filaments.

\begin{figure}[h]
\label{ContourFig}
\plotone{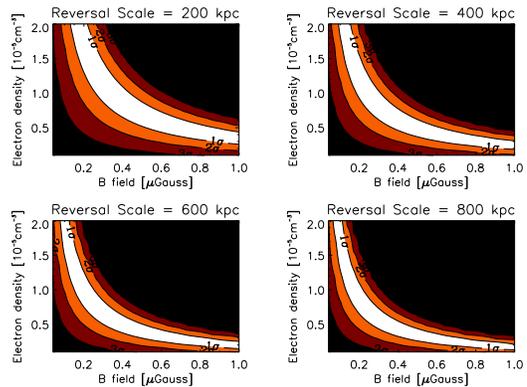}
\caption{Constraints on magnetic field and thermal electron density in
the Hercules supercluster for magnetic field reversal scales of 200, 400,
600, and 800~kpc respectively. The contours in each plot show the
$1\sigma$, the $2\sigma$, and the $3\sigma$ consistency contours
respectively.}
\end{figure}

\section{Summary and conclusion}
\label{ConcSec}

We have performed the first search for intergalactic magnetic fields
in three galaxy-overdense zones of the nearby universe: the Virgo,
Hercules, and the Perseus-Pisces superclusters of galaxies. We have
compared our results with simplified model expectations of IGM
magnetic fields and intergalactic thermal gas, the latter being
represented by the WHIM.  Algorithms were developed to calculate zones
of galaxy overdensity in these three areas using the 2MASS and CfA2
galaxy survey data, and combined with new Faraday rotation measures.

Our results are summarized as follows:

(1) The large angular extent of the Virgo supercluster combined with
its relatively low galaxy surface density make any Faraday RM
contribution difficult to distinguish from small variations in the
foreground RM of our Galaxy. We obtain a null result with the best
available data on RM and galaxy counts, which have inadequate
signal-to-noise for the purpose.

(2) For the Hercules supercluster, the Faraday RMs are enhanced
overall within the supercluster boundary on the sky. However within
the boundaries, the enhanced RMs do not generally occur precisely at
zones of maximum supercluster pathlength.  This may reflect the
existence of large scale coherence of extragalactic magnetic fields
within the Hercules Supercluster, but outside of the most
galaxy-overdense zones within the supercluster volume.

(3) The Perseus-Pisces supercluster, which coincides with a larger
``region A'', a zone of anomalously large RMs off the Galactic plane
(Simard-Normandin and Kronberg 1980), shows the clearest detailed
correlation with galaxy column density and supercluster volume
pathlength for background Faraday-rotated radio sources. This
correlation makes it likely, but not absolutely certain that some of
the enhanced RMs are actually generated in the IGM of the supercluster
over and above a Milky Way contribution. As with the Hercules
supercluster, future larger numbers of pulsar RMs are needed to
disentangle the relative RM contributions from the Milky Way and the
ICM of the Perseus-Pisces supercluster/filament zone.

(4) Comparison with simple models of ``expectation'' IGM fields gives
a result consistent with typical IGM fields in the supercluster IGM of
order 0.3 $L_{500\rm{kpc}}^{-0.5}$~$\mu$G at a $2\sigma$ significance
level.

Future advances in probing the Faraday Depths \citep{burn1966,
debruyn05} of supercluster environments will require (1) much denser
sky coverage of discrete source Faraday rotation measurements as
\citet{Gaensler2005} have recently achieved for the Magellanic clouds,
and (2) a better specification of the galactic magneto-ionic
environment at off-plane directions using deeper surveys of pulsar RMs
and dispersion measures, and multi-frequency all-sky radio
polarization surveys. In addition, we require (3) fainter
spectroscopic X-ray detections of intergalactic gas transitions to
better specify the temperature and density of the WHIM. The first two
will be easily accomplishable with the next radio telescope generation
such as the proposed ``Square Kilometer Array'' \citep{SKA04}. At
present, the SDSS and other low redshift optical galaxy surveys are
the most complete portion of the entire suite of observations required
to better characterize the physics of the IGM.

\section*{acknowledgments}

This publication makes use of data products from the Two Micron All
Sky Survey, which is a joint project of the University of
Massachusetts and the Infrared Processing and Analysis
Center/California Institute of Technology, funded by the National
Aeronautics and Space Administration and the National Science
Foundation. At Los Alamos National Laboratory, this research is
supported by the Department of Energy, under contract
W-7405-ENG-36. Support is also acknowledged (PPK and QWD) from the
Natural Sciences and Engineering Research Council of Canada.

 




\begin{thebibliography}{}

\bibitem[2MASS~Team(2003)]{2MASS03}
2MASS Team, 2003, http://www.ipac.caltech.edu/2mass/

\bibitem[Abazajian et al.(2003)]{aba03}
	Abazajian, K. et al. 2003, AJ, 126, 2018 (DR1 paper)

\bibitem[Abazajian et al.(2004)]{aba04}
	Abazajian, K. et al. 2004, AJ, 128, 502 (DR2 paper)

\bibitem[Abazajian et al.(2005)]{aba05}
	Abazajian, K. et al. 2005, AJ, 129, 1755 (DR3 paper)

\bibitem[Athreya et al. (1998)]{athre98}
	Athreya, R.M., Kapahi, V.K., McCarthy, P.J., \& van Breugel, W. 1998, \aap, 329, 809-820

\bibitem[Barber, Dobkin \& Huhdanpaa (1996)]{qhull96}
    Barber, C.B., Dobkin, D.P., \& Huhdanpaa, H. 1996, ACM
Trans. Math. Software 22, 469; http://www.qhull.org/

\bibitem[Biermann(1950)]{Biermann50}
	Biermann, L. 1950, Z. f\"{u}r Naturforschung A, 5, 65

\bibitem[Brentjens \& deBruyn(2005)]{debruyn05}
	Brentjens, M., \& deBruyn 2005, A\&A (in press) 

\bibitem[Burn(1966)]{burn1966} Burn, B.J. 1966, MNRAS, 133, 67

\bibitem[Carilli \& Rawlings(2004)]{SKA04}
	Carilli, C. \& Rawlings, S. (eds.) ``Science with the Square Kilometre Array",
	2004, New Astronomy, 48

\bibitem[CfA2~Team(1998)]{CfA98}
CfA2 Team, 1998, http://cfa-www.harvard.edu/$\sim$huchra/zcat/

\bibitem[Chokshi \& Turner(1992)]{Choks1992} 
Chokshi, A., \& Turner, E.L. 1992, MNRAS, 259, 421

\bibitem[Clarke \& Ensslin(2000)]{clarke2000}
Clarke, T.E. \& En{\ss}lin, T.A., 2005, \aj, in press 

\bibitem[Clarke, Kronberg \& B\"{o}hringer(2001)]{clarke2001} 
Clarke, T.E., Kronberg, P.P., \& B\"{o}hringer H. 2001, \apj, 547, L111

\bibitem[Colgate \& Li (2004)] {colgate2004} Colgate, S.A. \& Li,
H. 2004, {\it Comptes rendus - Physique}, ed. Sigl,~G. \&
Boratav,~M. {\bf 5}, 431 - 440

\bibitem[Dolag et al (2005)]{dolag2005} Dolag, K., Grasso, D.,
Springel, V., \& Tkachev, I. 2005, {\it Journal of Cosmology \&
Particle Phys.}, {\bf 1}, 9

\bibitem[Feretti et al.(1995)]{Feretti95} Feretti,~L., Dallacasa,~D.,
Giovannini,~G., \& Tagliani,~A. 1995, A\&A, 302, 680.

\bibitem[Furlanetto \& Loeb(2001)]{FL01} 
Furlanetto, S.R., \& Loeb, A. 2001, \apj, 556, 619

\bibitem[Gaensler et al.(2005)]{Gaensler2005}
Gaensler, B. M. et al. 2005, Science, 307, 1610-1612

\bibitem[Gebhardt et al.(2000)]{Gebh00} Gebhardt et al. 2000, \apj,
539, L13-16 

\bibitem[Gnedin, Ferrara \& Zweibel]{GFZ00} Gnedin, N.Y., Ferrara, A.,
\& Zweibel, E.G. 2000, \apj, 539, 505 

\bibitem[G\'{o}rski, Hivon, \& Wandelt(1999)]{GHW}
G\'{o}rski,~K.M., Hivon,~E., \& Wandelt,~B.D., in
Proceedings of the MPA/ESO Cosmology Conference ``Evolution of
Large-Scale Structure", eds. Banday,~A.J., Sheth,~R.S., and Da
Costa,~L., PrintPartners Ipskamp, NL, pp. 37-42 (also
astro-ph/9812350)  

\bibitem[Harrison(1970)]{Harrison70} Harrison,~E.R. 1970, MNRAS, 147,
279. 

\bibitem[Haslam et al.(1982)]{} Haslam, C.G.T, Stoffel, H., Salter, C.J, \& Wilson, W.E. 1982, 
\aaps, 47, 1    

\bibitem[Icke \& van de Weygaert(1987)]{Icke87}
	Icke, V. \& van de Weygaert, R. 1987, \aap, 184, 16

\bibitem[Kim et al.(1989)]{Kim1989} Kim, K.T., Kronberg, P.P.,
Giovannini, G., \& Venturi, T. 1989, Nature, 341, 720

\bibitem[Kim et al.(1990)]{kim1990} Kim, K.T., Kronberg, P.P.,
Dewdney, P.D., \& Landecker, T.L. 1990, \apj, 355, 29

\bibitem[Kim et al.(1991)]{kim1991} 
Kim, K.T., Tribble, P., \& Kronberg, P.P. 1991, \apj, 379, 80, 1991. 


\bibitem[Kronberg et al.(2001)]{Kronb2004} 
Kronberg, P.P., Colgate, S.A., Li, H., \& Dufton, Q.W. 2004, \apj,
604, L77 

\bibitem[Kronberg, Lesch, \& Hopp(1999)]{KLH1999} Kronberg,~P.P.,
Lesch,~H., \& Hopp,~U. 1999, ApJ, 511, 56.

\bibitem[Kronberg et al.(2001)]{kronb2001} 
 Kronberg, P.P., Dufton, Q.W., Li, H., \& Colgate, S.A. 2001, \apj,
560, 178 

\bibitem[Kronberg \& Perry(1982)]{kronb1982} Kronberg, P.P. \& Perry,
J.J. 1982, \apj, 263, 518  

\bibitem[Kronberg et al.(1986)]{kronb1986} Kronberg, P.P.,
Wielebinski, R., \& Graham, D.A. 1986, \aap, 169, 63

\bibitem[Mack et al.(1997)]{mack1997} Mack, K.-H., Klein, U., O'Dea, C.P., \& Willis, A.G. 1997, \aaps, 123, 423-444

\bibitem[Nicastro et al.(2005a)]{Nic2005a}
Nicastro, F., et al. 2005a, Nature, Volume 433, Issue 7025, pp. 495-498

\bibitem[Nicastro et al.(2005b)]{Nic2005b}
Nicastro, F., et al. 2005b, astro-ph/0501126

\bibitem[Ryu et al (1998)]{ryu98}
Ryu, D.S., Kang, H., \& Biermann, P.L. 1998, \aap, 335, 19 - 25

\bibitem[Schlegel, Finkbeiner \& Davis(1998)]{SFD98}
Schlegel, D. J., Finkbeiner, D. P., \& Davis, M. 1998, \apj, 500, 525

\bibitem[Simard-Normandin \& Kronberg(1980)]{SK80}
Simard-Normandin, M. \& Kronberg, P. P. 1980, \apj, 242, 74-94

\bibitem[Simard-Normandin, Kronberg \& Button(1981)]{SKB81}
Simard-Normandin, M., Kronberg, P.P., \& Button, S. 1981 \apj, Suppl.,
45, 97-111 

\bibitem[Small \& Blandford(1992)]{small1992} Small, T.A., \&
Blandford, R.D. 1992, MNRAS 259, 725 

\bibitem[Soltan(1982)]{Soltan82} Soltan, A. 1982, MNRAS, 200, 115

\bibitem[Strom \& Willis(1980)]{strom80}
	Strom, R.G., \& Willis, A.G. 1980, \aap, 85, 36


\bibitem[Welter et al.(1984)]{welt1984} Welter, G.L., Perry, J.J. \&
Kronberg, P.P. 1984, \apj, 279, 19 

\bibitem[Willis \& Strom(1978)]{willi1978} Willis, A.G., and Strom,
R.G. 1978, \aap, 62, 375

\bibitem[Xu(2005)]{xu2005} Xu, Y. 2005, in preparation

\end{thebibliography}
\end{document}